\documentstyle[aps]{revtex}
\input{psfig}
\begin{document}
\title
{ 
 \Large\bf On the Parity Degeneracy of Baryons 
 }
\author {
M. Kirchbach \\
{\small \it Institut f\"ur Kernphysik, J. Gutenberg Universit\"at, 
  D-55099 Mainz, Germany}
}
\maketitle

\begin{abstract}
The gross features of the observed baryon excitation 
spectrum below 2 GeV are well explained if the spectrum generating 
algebra of its intrinsic orbital angular momentum states 
is o(4)$\otimes$su(2)$_I$. The spins of the resonances are 
obtained through the coupling of a Lorentz bi--spinor 
$\lbrace {1\over 2} ,0\rbrace \oplus \lbrace 0, {1\over 2} \rbrace $
to a multiplet of the type $\lbrace j,j \rbrace $ in its O(4)/O(3) 
reduction. The parities of the resonances follow from those of the
O(3) members of the $\lbrace j,j\rbrace $ multiplets.
In this way relativistic SL(2,C) representations are constructed.
{}For example, the first $S_{11}$, $P_{11}$, and D$_{13}$ 
states with masses around 1500 MeV fit into the 
$\lbrace {1\over 2}, {1\over 2} \rbrace \otimes 
\lbrack \lbrace {1\over 2},0\rbrace 
\oplus \lbrace 0,{1\over 2} \rbrace\rbrack $ representation. 
The observed parities of the resonances correspond to natural
parities of the $\lbrace {1\over 2}, {1\over 2} \rbrace $ states. 
The second $P_{11}$, $S_{11}$, $D_{13}$ -- together with
the first $P_{13}$, $F_{15}$, $D_{15}$, and (a predicted)
$F_{17}$--resonances, centered around 1700 MeV, are organized
into the $\lbrace {3\over 2},{3\over 2} \rbrace \otimes\lbrack 
\lbrace {1\over 2},0\rbrace \oplus \lbrace 0,{1\over 2}\rbrace
\rbrack $ representation. I argue that the members of the 
$\lbrace {3\over 2}, {3\over 2}\rbrace $ multiplet carry unnatural 
parities and that in this region chiral symmetry is restored. 
In the $N(939)\to N(1650)$ transition the chiral symmetry mode
is changed, and therefore, a chiral phase transition is predicted
to take place.

KEY  words:  parity doublets, baryon resonances,
            chiral symmetry restoration, spinors.

\end{abstract}

\newpage

\section{ Introduction}
The occurrence of more or less mass degenerate states of equal spins
and opposite parities in both the baryonic and the mesonic spectra
 \cite{PDGr} has been subject of only few theoretical 
investigations. For example, while the nucleon and its 
first spin ${3\over 2}^-$--resonance (D$_{13}$ state) 
appear as singlets, the low lying spin--${1\over 2}$ resonances 
(the first and second S$_{11}$ and P$_{11}$ states) of opposite 
parities are almost mass degenerate. 
One possible explanation of such phenomena was given by Dashen in the 
late sixties \cite{Dashen}. In this work, the baryon spectrum was
classified according to the representations of the group
SU(3)$_f \otimes {\cal Z}$, the direct product of the special three 
flavor symmetry group, and a discrete group ${\cal Z}$, constituted 
by the six elements $\lbrace $ 1, ${\cal P}$, Z:= exp($i{{2\pi}\over 3} 
\lambda_8\gamma_5 )$,
Z$^\dagger $, ${\cal P}$Z, ${\cal P}$Z$^\dagger \rbrace $ with ${\cal P}$ 
being the 3--space inversion operator.
As long as the group ${\cal Z}$ has two one--dimensional representations 
that are parity singlets, and a further two--dimensional 
(parity degenerate) representation, the baryons can be organized into 
parity singlets on the one side, and parity doublets on the other side. 
A parity doublet is then characterized by the small couplings 
(denoted by $g_{\pi N}^{L_{2I,2J}}$) of comparable size 
of mass degenerate resonances of opposite parities 
to the pion--nucleon ($\pi $N) channel containing singlets only. 
Within this model, the conclusion was made that 
whereas the first S$_{11}$ and P$_{11}$ states might constitute a 
parity doublet, the mass degeneracy of the first F$_{15}$ and D$_{15}$ 
resonances might rather be accidental. 

In the light of contemporary knowledge, the model of 
Ref.~\cite{Dashen} needs further specifications, basically because of 
the different behavior of $g_{\pi N}^{S_{11}} $ and $g_{\pi N}^{P_{11}} $
in the large $N_c$ limit. Indeed, it was shown in \cite{WiKi} that while 
$g_{\pi N}^{P_{11}} $ is of the order ${\cal O}(N_c)$, 
 $g_{\pi N}^{S_{11}}$ is of the order ${\cal O}(1/N_c)$
thus revealing more differences between the $S_{11}$ and P$_{11}$ 
resonances than naively expected from a parity doublet classification.
Later, in accordance with an idea advocated by Iachello  \cite{Franko},
the existence of mass degenerate resonances of equal spins and
opposite parities was attributed to the 
geometric structure of baryons, suggested to be supplementary to 
their internal color and spin--flavor structures.
The assumption of an asymmetric top shape of hadrons led then to
a classification of baryons according to 
the geometric symmetry group $C_{3v}$ at low and  $C_{\infty v}$
and at high angular momenta, respectively, and 
to the $C_{\infty v }$ symmetry for the mesonic $\bar q q$ string. 
An essential conclusion of this model is that only for 
an ideally spherical top shape parity duplication is exact.
Indeed, the symmetry group O(3) of such a top can be represented as a 
direct product of the special orthogonal group in three dimensions SO(3) 
and the discrete group S$_2$ constituted 
by the identity and the space inversion elements. Since S$_2$ has two 
one--dimensional representations corresponding to opposite parities, 
the SO(3) multiplets of angular momentum $l$ split into natural 
($(-1)^l$) and unnatural ($(-1)^{l+1}$) parity states in the full 
group O(3). In case the hadron shape becomes  deformed from a 
spherically symmetric to, say, an octupol one, then the geometric 
symmetry group changes from O(3) to  $C_{\infty v}$ and the parity 
duplication is no longer exact. 

This latter group is well known from solid state physics 
as the limit $n\to \infty $ (so called Curie limit) of the point 
group $C_{ n v} = C_n\otimes \sigma_v$. For example, $C_3\otimes 
\sigma_v $ is the symmetry group of CH$_3$Cl where $C_3$ is the 
trigonal point group and $\sigma_v$ the
reflection with respect to a plane containing the Cl-C molecule axis. 
It is well known from group theory (compare \cite{EllDaw} for details) 
that among the irreducible representations of $C_{\infty v }$  there is 
a band of natural parity states appearing as singlets as well as 
degenerate multiplets. A comparison of the experimental baryon spectrum 
with this geometric nucleon model reveals, however, 
that some superfluous parity doublets are predicted already 
below 2 GeV, for which no identification with physical states can be 
made. Finally, both treatments of the baryon spectrum ~\cite{Dashen} 
and \cite{Franko} are based on nonrelativistic group symmetries and do 
not lead to relativistic equations of motion for baryons with higher 
spin. For these reasons it appears timely to re--examine the parity 
doublet structure of the baryonic spectrum.

In the present paper I demonstrate that the baryon spectrum below 
2 GeV is well explained if the spectrum generating algebra of the 
intrinsic orbital angular momentum of the resonances is 
o(4)$\otimes$ su(2)$_I$, i.e., the direct product of the compact 
orthogonal group in four Euclidean dimensions and the isospin group. 
The paper is organized as follows.
In the next section I briefly review the properties of the
group SO(4), introduce its low lying irreducible representations,
and perform the classification of the orbital angular momenta,
underlying the baryon resonances below 2 GeV in terms of SO(4). 
In Sect. III, I briefly sketch the development of Lorentz 
invariant equations of motion for particles of any spin. 
I advocate the viewpoint that the relativistic equation of motion
for the  resonances has to be concluded from the projector onto the
SL(2,C) representation embedding these particles in the respective 
excitation spectrum. From that I conclude that the relativistic 
equation of motion for the baryonic $\lbrace {1\over 2}, 
{1\over 2}\rbrace  \otimes \lbrack \lbrace {1\over 2},0 \rbrace 
\oplus \lbrace 0, {1\over 2}\rbrace \rbrack $ multiplet known as the 
Rarita--Schwinger equation describes a family of the states 
P$_{2I\, ,1}$, S$_{2I\, ,1}$ and D$_{2I\, ,3}$ (or P$_{2I\, , 3}$)  
as a whole rather than, as widely used, an isolated P$_{2I\, , 3}$ 
state. The paper ends with a short summary.

\section{ O(4)$\otimes $SU(2)$_I$ classification of the baryon 
          resonances}

The group O(4) is the direct product of the special orthogonal
group SO(4) and the discrete group S$_2$ mentioned in the
introduction. There are two different types of O(4) multiplets.
In the first case the O(3) members of the O(4) multiplets 
are of natural parities, while in the second case their parities are 
unnatural. Subsequently, the SO(4) representations will be considered 
in more detail.

The group SO(4) is well known to be the factor group of the chiral group
SU(2)$_L\otimes $SU(2)$_R$ as the chiral algebra $su(2)_L\oplus su(2)_R$
is the universal enveloping of $so(4)$. Because of that, the 
representations of the chiral group can be mapped onto those of SO(4).
For this reason SO(4) transformations are sometimes called 'chiral'.  
Along this line, the famous SO(4) invariant mesonic lagrangian known as 
the linear sigma model had been constructed by Gell--Mann and Levy 
\cite{GLevy}. To my knowledge, no SO(4) invariant models for baryon 
excitations have been considered so far. It is the purpose of the present 
study to partly fill this gap.

The special orthogonal group in four dimensions is the
compact form of the special Lorentz group SO(1,3). To be more general,
I present here the commutation relations between 
the six generators $G_i$ and $\widetilde{G}_i$ with i=1,2,3 
of the non--compact group SO(1,3). They generate the following algebra:
\begin{eqnarray}
\lbrack G_i,G_j\rbrack = i\epsilon_{ijl}G_l\, ,   \quad
\lbrack \widetilde{G}_i, \widetilde{G}_j\rbrack = -i\epsilon_{ijl}G_l\, ,
&&
\lbrack G_i, \widetilde{G}_j\rbrack = i \epsilon_{ijl}\widetilde{G}_l\, , 
\nonumber\\
\lbrack \widetilde{G}_i, G_j\rbrack = i \epsilon_{ijl}\widetilde{G}_l\, ,
&\quad &
\lbrack G_k, \widetilde{G}_k\rbrack =  0\,, \quad\quad  i,j,k= 1,2,3 \, . 
\label{so4_alg}
\end{eqnarray}
The $so(4)$ algebra is simply obtained from the last equation
through the replacement $\widetilde{G}_k \to i \widetilde{G}_k$. This 
complexification of the so(1,3) algebra is sometimes called 
'Wick rotation'. The irreducible representations $\lbrace j, j' \rbrace $ 
of the group SO(4) are labeled by the two numbers $j$ and $j'$ that are 
both integer or both half--integer, and their dimensionality is 
\begin{equation}
{\tt dim}\lbrace j, j'\rbrace  = (2j +1) (2j' +1)\, .
\label{dim_so4}
\end{equation}
The SO(4) multiplets have in the SO(3) reduction the following
angular momentum content 
\begin{equation}
l= \mid j - j' \mid , ... , j + j'\, .
\label{l_so3}
\end{equation}
The irreducible representations $\lbrace l\rbrace $ of SO(3) split within 
the full group O(3) into states of natural and unnatural parities. 
The natural parity (n.p.) states are realized in terms the spherical 
harmonics Y$^l_m (\hat{r})$, while the unnatural parity (un.p.) states are 
described by means of spherical pseudotensors. 

In the full group O(3), natural and unnatural 
parity states are distinguishable, while in the special group they are 
identical. For example, the lowest SO(4) multiplet  $\lbrace 0,0 \rbrace $ 
is one--dimensional and remains a singlet after the SO(3) reduction. Its  
parity and angular momentum content with respect to O(3) can be 0$^+$ as 
well as 0$^-$. The manifest chiral symmetry O(4)$\otimes $SU(2)$_I$ 
suggests, therefore, that all states of natural parity are paralleled by 
such of unnatural parity. On the other hand,
its spontaneous breaking implies the selection of the 0$^+$ state and thus 
of the natural parity for the lowest O(3) representations. 
The next higher dimensional O(4) representation is the one with 
$j = j' = {1\over 2} $. This is a 4--dimensional multiplet and its orbital 
angular momentum content in the O(4)/O(3) reduction can be either of natural 
or unnatural parity according to
\begin{eqnarray}
\lbrace {1\over 2}, {1\over 2}\rbrace 
&\stackrel{O(3)_{n.p.}}{\longrightarrow}& 0^+\oplus 1^-\, ,\\
\lbrace {1\over 2}, {1\over 2}\rbrace 
&\stackrel{O(3)_{un.p.}}{\longrightarrow}& 0^-\oplus 1^+\, .
\label{4vector_so4}
\end{eqnarray}

For the purpose of the present investigation I wish to introduce also
two more O(4) representations that will play a role in the following.
These are the 16--plet $\lbrace {3\over 2}, {3\over 2}\rbrace $,
and the 36--plet $\lbrace {5\over 2},{5\over 2}\rbrace $ for which one 
finds the following O(3) reductions
\begin{eqnarray}
\lbrace {3\over 2}, {3\over 2}\rbrace &\stackrel{O(3)_{n.p.}}
{\longrightarrow}& 0^+\oplus 1^-\oplus 2^+ \oplus 3^-\,,\\
\lbrace {3\over 2}, {3\over 2}\rbrace &\stackrel{O(3)_{un.p.}}
{\longrightarrow}& 0^-\oplus 1^+\oplus 2^-\oplus 3^+\, ,
\label{16plet_so4}\\
\lbrace {5\over 2}, {5\over 2}\rbrace &\stackrel{O(3)_{n.p.}}
{\longrightarrow}& 0^+\oplus 1^-\oplus 2^+ \oplus 3^-\oplus 4^+
\oplus 5^- \, ,\\
\lbrace {5\over 2}, {5\over 2}\rbrace &\stackrel{O(3)_{un.p.}}
{\longrightarrow}& 0^-\oplus 1^+\oplus 2^- \oplus 3^+\oplus 4^-
\oplus 5^+\, . \label{36plet_so4}
\end{eqnarray}
Now, one can consider the total angular momentum $\vec{J}$
of the baryon resonances to be generated through 
the coupling of a spin--${1\over 2}$ state to the O(4) representation 
$\lbrace j, j\rbrace $ filled with the 
corresponding orbital angular momentum states $l=0,..., 2j$ of 
natural/unnatural parities in accordance with 
$J=|l-{1\over 2}|,(l+{1\over 2})$. 
Similar type of couplings appear, for example, 
in the construction of the wave functions for the excited states of the 
hydrogen atom. There, the spin of an electron moving within the SO(4) 
symmetric Coulomb potential of the proton is coupled to the 
corresponding Coulomb multiplet with main quantum number 'n' via 
with $|\vec{j}|=|l-{1\over 2}|, (l+{1\over 2})$ 
and $l$=0,1,..., n-1. The parity of the hydrogen states 
obtained in this way is then determined by $(-1)^l$.
In the baryon case one might imagine instead of the
electron--proton system a valence quark with a di--quark in its 
ground state. 

The parity $(-1)^{L+1}$ of a single $\pi N$ resonance
$L_{2I \, , 2J }$ (in the standard notation of \cite{PDGr} )
is then equivalently determined in the present classification 
by either $(-1)^l$ or $(-1)^{l+1}$ depending on whether the parity of 
the O(3) state is natural or unnatural. 
In the present notation, $L$ takes the values of either
$L= | l-1 |,(l+1)$ for natural, or $L=l$ for unnatural parities.
For example, while a $\lbrace {1\over 2}, {1\over 2}\rbrace $
multiplet of natural parity is filled with P$_{2I\, , 1}$, S$_{2I\, ,1}$, 
and D$_{2I\, ,3}$ resonances, the one filled with unnatural parities
would require the S$_{2I\, , 1}$, P$_{2I\, , 1}$, and P$_{2I\, , 3}$ 
states. This consideration illustrates that the parity of the highest 
spin resonance allows to reconstruct the parity of the O(3) members of 
the $\lbrace j,j\rbrace $ multiplet. 

For completeness it should be noted that
the spin--${1\over 2}$ state has to be taken as the (reducible) 
representation $\lbrace {1\over 2}, 0 \rbrace \oplus 
\lbrace 0, {1\over 2} \rbrace $ of the group SL(2,C) acting as the 
universal covering of the special Lorentz group SO(1,3), the compact form 
of which is just SO(4). Note, that combining spinor 
$\lbrace {1\over 2}, 0\rbrace $ and 
co--spinor $\lbrace 0, {1\over 2}\rbrace $ to a bi--spinor enables one to 
cover the space-- and time reflection transformations of the full Lorentz 
group O(1,3) \cite{Scheck,RumFet}. Within this framework, the nucleon is 
obviously classified according to the relativistic multiplet
\begin{equation}
{1\over 2} \left({1\over 2}^+ \right) \simeq 
               \lbrace 0,0 \rbrace \otimes \lbrack
               \lbrace {1\over 2},0 \rbrace
\oplus \lbrace 0, {1\over 2 }\rbrace \rbrack \otimes \lbrace   
\mbox{ \boldmath${1\over 2}$ }\,\rbrace  , 
\label{nucleon}
\end{equation}
while the first excited P$_{11}$, S$_{11}$ and D$_{13}$ states with 
masses 1440 MeV, 1535 MeV, and 1520 MeV, respectively, 
are organized into
\begin{equation}
{1\over 2} \left(
{1\over 2}^+, {1\over 2}^-, {3\over 2}^- \right)\simeq
\lbrace {1\over 2},{1\over 2}\rbrace \otimes 
\lbrack \lbrace {1\over 2},0\rbrace \oplus
\lbrace 0, {1\over 2} \rbrace \rbrack \,\otimes 
\lbrace \mbox{ \boldmath${1\over 2}$ }\rbrace  \,  .
\label{Roper_group}
\end{equation}
The notation  $I(J^\pi )$ on the lhs in Eqs.~(\ref{nucleon}), and 
(\ref{Roper_group}) denotes a resonance with isospin $I$, and spin and 
parity $J^\pi$, respectively, while on the rhs the isospin of the
multiplet is denoted by $\lbrace  \mbox{ \boldmath$I$ }\rbrace $.
Furthermore, $'\simeq ' $ was used to denote the mapping of the baryon 
states onto SL(2,C) representations. 

The second P$_{11}$, S$_{11}$ and D$_{13}$ states 
with masses of 1710 MeV, 1650 MeV, and 1700 MeV, respectively,
together with the lowest P$_{13}$, F$_{15}$,  D$_{15}$, and a (still 
unobserved) G$_{17}$ (natural parity) or 
F$_{17}$ (unnatural parity) resonances with masses between 
1675 MeV and 1720 MeV can well be joined together into the relativistic 
SL(2,C) multiplet
\begin{equation}
{1\over 2}\left(
 {1\over 2}^\pm, {1\over 2}^\mp, {3\over 2}^\mp, {3\over 2}^\pm,
{5\over 2}^\pm , {5\over 2}^\mp, {7\over 2}^\mp, \right) \simeq
\lbrace {3\over 2}, {3\over 2} \rbrace\otimes 
\lbrack \lbrace {1\over2}, 0\rbrace \oplus 
\lbrace 0, {1\over 2} \rbrace\rbrack  \,  \otimes 
\lbrace \mbox{ \boldmath${1\over 2}$ } \rbrace  \, ,
\label{second_group}
\end{equation}
where the upper/lower sign corresponds to natural/unnatural parity
of the intrinsic angular momentum states.
The parity of  the spin--${7\over 2}$ resonance reveals
whether or not chiral symmetry is restored at this energy. 
Indeed, in case of a G$_{17}$ state the O(4) 16--plet 
would still belong to the Fock space built upon the spontaneously
selected vacuum of natural parity. In such a case, 
chiral symmetry is still in the Nambu--Goldstone mode.
On the contrary, the case of a F$_{17}$ state would reveal 
the appearance of unnatural parity states and thus signal 
the Wigner-Weyl mode of chiral symmetry. In order to make a prediction 
for the parity of the first spin--${7\over 2}$ nucleon excitation, a 
comparison of the $I={1\over 2}\, (N)$ and $I={3\over 2}\, (\Delta )$ 
spectrum is quite instructive. 

It is easily proven that the classification scheme developed so far for 
$I={1\over 2}$ states, applies equally well to $\Delta $ 
resonances. First of all, each one 
of the two lowest P$_{33}$ states at 1232 MeV and 1600 MeV 
is attached to a single representation 
\begin{equation}
{3\over 2}\left({3\over 2}^+\right) \simeq \lbrace 0,0\rbrace \otimes 
\lbrack
\lbrace {3\over 2}, 0\rbrace \oplus \lbrace 0, {3\over 2} \rbrace\rbrack 
\, \otimes \lbrace \mbox{\boldmath${3\over 2}$}\rbrace \,  .
\label{lowest_Deltas}
\end{equation}
The appearance of the second P$_{33}$ state is of dynamical
origin. For example, such a state could contain hybrid nucleon--gluon 
(Ng) components \cite{BaCl}.
Note that a second $\lbrace {1\over 2}, 0\rbrace \oplus 
\lbrace 0,{1\over 2} \rbrace $ state of positive parity  is 
still missing in the nucleon spectrum.
Furthermore, the P$_{31}$, S$_{31}$ and D$_{33}$ with masses 1750 MeV, 
1620 MeV, and 1700 MeV, respectively, fit into 
\begin{equation}
\lbrace 
{1\over 2}, {1\over 2}\rbrace \otimes\lbrack
\lbrace {1\over 2}, 0\rbrace \oplus \lbrace 0, {1\over 2}\rbrace 
\rbrack \,\otimes 
\lbrace  \mbox{ \boldmath${3\over 2}$ }\rbrace \, ,
\label{Delta_res}
\end{equation}  
and all possess natural parities.

The next isoquadruplet excitations P$_{31}$, S$_{31}$, D$_{33}$,
together with P$_{33}$, F$_{35}$, D$_{35}$ and the F$_{37}$ 
resonances with masses ranging from 1900 MeV to 1950 MeV are well 
organized into the O(4) 16-plet of \underline{unnatural} parity
\begin{equation}
{3\over 2}({1\over 2}^-, {1\over 2}^+,{3\over 2}^+,
{3\over 2}^-,{5\over 2}^-, {5\over 2}^+, {7\over 2}^+)\simeq
\lbrace {3\over 2}, {3\over 2}
\rbrace \otimes \lbrack
\lbrace {1\over 2},0 \rbrace
\oplus \lbrace 0, {1\over 2}\rbrace \rbrack \, \otimes
\lbrace  \mbox{ \boldmath${3\over 2}$ }\rbrace \, ,
\label{higher_deltas}
\end{equation}
where they  are obtained in coupling a spin-${1\over 2}$ state 
with orbital angular momenta $0^-,1^+,2^-$, and $3^+$, 
respectively (see Table 1). The positive parity of the 
spin--${7\over 2}$ $\Delta $ state indicates unnatural parities 
for the states of the 16--plet. Now, it seems natural to assume 
the existence of a F$_{17}$ state that parallels the F$_{37}$ 
resonance and thereby to predict unnatural parities for the states 
of the nucleon 16--plet, too. In doing so, the second nucleon 
S$_{11}$ resonance at 1650 MeV acquires importance as a 'would-be' 
parity partner of the nucleon.

The notion of a 'would be' parity partner to the nucleon for a 
S$_{11}$ state was first introduced in \cite{DeTar} in connection 
with the parity degeneracy of baryons above the critical
temperature for chiral transitions following from lattice QCD
calculations of the hadron screening length.  
There, the nucleon--S$_{11}$ mass splitting was generated through the 
spontaneous chiral symmetry breaking within the framework of an 
appropriately extended $\sigma $ model. Within such a scenario 
(see \cite{HaPr} and references therein) the nucleon--S$_{11}$ 
transition was predicted to be a first order chiral phase transition. 
In \cite{HaPr} it was the first S$_{11}$ state that was considered
as the 'would-be' parity partner of the nucleon. In using
the width, mass and coupling constant to the $\pi N$ channel
of that state, the initial ($\rho_i$) and final ($\rho_f$) densities 
of the chiral phase transition in nuclear matter
where predicted as 3.1$\rho_0$ and 6.4$\rho_0$, with
$\rho_0$ being the normal matter density, respectively.

From the SL(2,C) baryon classification follows that the first S$_{11}$
resonance can not be viewed as the parity partner of
the nucleon because it belongs to a relativistic representation built
upon the spontaneously selected vacuum of natural parity.
Following the arguments given above, the baryon spectrum below
1.6 GeV is still in the Nambu--Goldstone mode of chiral symmetry.
On the contrary, the second S$_{11}$ state was shown to
belong to a relativistic multiplet built upon a vacuum state of 
unnatural parity. For that reason, the parameters of the N(1650) 
state have to enter the calculation of the critical matter density of
the chiral phase transition for baryons. Since the second
S$_{11}$ is much stronger coupled to the $\pi N$ system than the
first one, the resulting changes of $\rho_i$ and $\rho_f$
might be substantial.  

As long as the classification of the baryon resonances according to 
the above SL(2,C) representations is the only way to understand the 
evident concentration of states within comparatively narrow mass 
regions in relation to their spin and parities, the chiral symmetry 
restoration for nucleons is expected to take place in the region 
around 1700 MeV. Correspondingly, for $\Delta $'s the chiral phase 
transition will take place at about 1900 MeV.  

Finally, the third P$_{11}$, S$_{11}$ and D$_{13}$ states with 
masses of 2100 MeV, 2090 MeV and 2080 MeV, respectively, the second 
P$_{13}$, F$_{15}$,  D$_{15}$, and G$_{17}$ resonances
in turn placed at 1900 MeV, 2000 MeV, 2200 MeV, and 2190 MeV  
together with the first F$_{17}$, H$_{19}$, G$_{19}$
and a further (still unobserved) spin--${11\over 2}$ resonances 
with masses ranging from 1990 MeV to 2250 MeV 
can be joined together into the following representation
\begin{equation}
{1\over 2}\left(
 {1\over 2}^\pm, {1\over 2}^\mp, {3\over 2}^\mp, {3\over 2}^\pm,
{5\over 2}^\pm , {5\over 2}^\mp, {7\over 2}^\mp, 
{7\over 2}^\pm,{9\over 2}^\pm, {9\over 2}^\mp, 
{{11}\over 2}^\mp \right) \simeq
\lbrace {5\over 2}, {5\over 2} \rbrace\otimes 
\lbrack \lbrace {1\over2}, 0\rbrace \oplus 
\lbrace 0, {1\over 2} \rbrace\rbrack \,  \otimes 
\lbrace \mbox{ \boldmath${1\over 2}$ } \rbrace \, .
\label{third_group}
\end{equation}
Thus a new spin--${{11}\over 2}$ state is predicted near 2200 MeV.
In order to make a prediction for the parity of the latter state, 
again a comparison between the $I={1\over 2}\, (N)$ 
and $I={3\over 2}\, (\Delta )$ 36--plets is useful. 
The full baryon particle listing in Ref.~\cite{PDGr} 
shows, unfortunately, that the $\Delta $ states from that mass 
region are still less known. Nonetheless, there is a well 
established H$_{3, 11}$ state at 2420 MeV that indicates that the 
observed 36--plet is filled by O(3) states of unnatural parity.
Therfore, it is natural to assume the spin--${{11}\over 2}$
nucleon resonance to be a H$_{1, 11}$ state, 
thus confirming the chiral symmetry restoration above 1.6 GeV.      
Furthermore, the SL(2,C) classification scheme predicts 
three new isoquadruplet states above 2000 MeV. These 
are the P$_{31}$, P$_{33}$, and D$_{33}$ state (see Table 2).

Comparison of the empirical baryon spectrum  \cite{PDGr}
with the above classification scheme shows that
all known non--strange baryon states below (and around) 2 GeV 
are well organized into SL(2,C) representations. The maximal mass 
splitting between two states within such a multiplet is about 90 MeV 
in the first nucleon resonance family, and about 70 MeV in the second 
family. For the $\Delta $ resonances these splittings are 130 MeV and 
50 MeV, respectively. The splitting within the SL(2,C) multiplets is 
significantly smaller relative to the gaps of about 200 MeV 
between the averaged masses of the resonance groups and might be 
attributed to spin dependent forces. Above 2 GeV, the SL(2,C) 
classification scheme seems violated more strongly with increasing mass 
and angular momentum. Within the third nucleon and $\Delta $ resonance 
families the mass splitting reaches 350 MeV, and 270 MeV,
respectively. Despite of that, these resonances form also  
well pronounced multiplets. 

I would like to note in passing that the classification scheme 
developed in the present work applies also to the $\Lambda $ hyperon 
spectrum (see Fig. 1). There, the isoscalar S$_{01}$, D$_{03}$, and 
P$_{01}$ states with masses from 1400 to 1600  MeV fit into the
$\lbrace {1\over 2}, {1\over 2}\rbrace \otimes
\lbrack \lbrace {1\over 2}, 0 \rbrace \oplus\lbrace 0, {1\over 2}
\rbrace \rbrack $ isosinglet representation. 
However, the mass splitting within the multiplet members is 
about two times larger as compared to the non--strange baryon sector.
Above 1600 MeV, the excitation spectrum of the $\Lambda $ hyperon
follows the pattern of the $C_{\infty v}$ symmetry of 
Ref. \cite{Franko} rather than those of SL(2,C).

\section{ Equations of motion for baryon resonances} 

All SL(2,C) representations constructed in Eqs.~(\ref{Roper_group}),
(\ref{second_group}), and (\ref{third_group}) are reducible. 
As long as the projectors onto these representations are well
known, writing relativistic equations of motion for the nucleon and 
its excited states becomes a well defined mathematical scheme. 
Examples for this are given in the present section.

The representation theory of the Lorentz group \cite{Scheck}
is a well established mathematical tool for obtaining
equations of motion for particles of any spin.
First of all one has to find in the rest frame the representation 
to which the particle under consideration belongs. 
To be specific, let us consider a $\lbrace j, j'\rbrace $ 
multiplet. The representation $\lbrace {1\over 2}, 0\rbrace$ with 
$j = {1\over 2}$ and $j' =0$ defines the spinor 
$\zeta = {\xi^1\choose \xi^2}$ while $\lbrace 0, {1\over 2} \rbrace $ 
introduces the co--spinor 
$\dot{\eta }={\eta_{\dot{1}} \choose \eta_{\dot{2}} }$
 \footnote{Note that the SL(2,C) group is known to have inequivalent 
representations in spinor-- and co--spinor spaces 
that can serve for describing particles and antiparticles.}.  
Particles with higher spin are then described in terms of spinor 
tensors ($S^{\mu \nu ...}_{\dot{\lambda} \dot{\beta }... }$)
of higher rank. For example, the spinor content of the vector
representation $\lbrace {1\over 2}, {1\over 2} \rbrace $ is 
$S^{\alpha \dot{\beta}} =\xi^\alpha \xi^{\dot{\beta }}$
with $\alpha =1,2$ and $\dot{\beta} = \dot{1}, \dot{2}$.
Especially, one gets the following correspondences
\begin{eqnarray}
x_0 + x_3 = \xi^1\xi^{\dot{1}}\, , &\quad & 
x_1 -ix_2 = \xi^1\xi^{\dot{2}}\, , \nonumber\\
x_1 +ix_2 = \xi^2\xi^{\dot{1}}\, ,  &\quad &
x_0 - x_3 = \xi^2\xi^{\dot{2}}\, ,
\label{quaternion}
\end{eqnarray}
where $\xi^1$ and $\xi ^2$ stand for the components of the 
spinor $\zeta $ while $\xi^{\dot{1}}$ and $\xi^{\dot{2}}$ are the 
respective components of the complex conjugate spinor.
Having fixed the representation of interest, one has to find 
the (2$j$+1)$\otimes $(2$j'$+1) matrices 
$ D^{\lbrace j,j'\rbrace }(b(\vec{p}\, ) )$ that carry out the 
boost transformation $b(\vec{p}\, )$ from the rest frame to the
system where the particle is moving with momentum $\vec{p}$.

Finally, the matrix representation of the 3--space inversion ${\cal P}$ 
has to be incorporated too. For example, to describe the space--inversion, 
$x_0 \to x_0$, and $\vec{x}\, \to - \vec{x}\, $ in Eq.~(\ref{quaternion}), 
the spinors must transform into co--spinors by means of an off--diagonal 
unit matrix denoted by $\gamma_0$ according to:
\begin{equation}
\left(
\begin{array}{c}
\zeta\\
\dot{\eta}\\
\end{array}
\right)
\stackrel{{\cal P}}{\longrightarrow}
\left(
\begin{array}{cc}
0&1\\
1&0
\end{array}\right)\left( 
\begin{array}{c}
\zeta\\
\dot{\eta}
\end{array}\right)\, := \gamma_0\left(
\begin{array}{c}
\zeta\\
\dot{\eta}
\end{array}\right)\, .
\label{reflection}
\end{equation}
The latter equation shows that the reflection operation in 3--space  
becomes possible only within the direct sum of the spinor-- and 
co--spinor spaces. All together,
the projector onto $\lbrace j, j'\rbrace $ is expressed as
\begin{equation}
\Pi (p)^{\lbrace j, j'\rbrace } =
D^{\lbrace j'j'\rbrace }(b(\vec{p}\, ) \,)
{1\over 2} (\eta_p {\cal P} +1)
(D^{\lbrace j,j'\rbrace }(b(\vec{p}\, ))^{-1}\, ,
\label{projector}
\end{equation}   
where $\eta_p$ is the parity of the vacuum.
As is seen from the non--diagonal form of the matrix $\gamma_0$
in Eq.~(\ref{reflection}), the spinors are mixed--parity states.
Because of that, the multiplets of the Lorentz group are often called 
'helicity' representations. Usually, one prefers to work within
the 'parity' representation where the matrix $\gamma_0$ is
diagonal. This is possible in the Pauli--Dirac  4--component basis
defined as
\begin{equation}
\Psi_1 = \xi^1 +\eta_{\dot{1}}\, , \quad 
\Psi_2 = \xi^2 +\eta_{\dot{2}}\, , \quad
\Psi_3 = -\xi^1 + \eta_{\dot{1}}\, , \quad
\Psi_4 = -\xi^2 + \eta_{\dot{2}}\, .
\label{parity_rpr}
\end{equation}
Now the projector onto spin--${1\over 2}$ particles for a 
vacuum state of positive parity $\eta _p =1$ and ${\cal P} =\gamma_0$ 
reads \cite{RumFet}
\begin{equation}
\Pi^D (p)  = {1\over {2m}}(/\!\!\!p +m)\, .
\label{dirac_proj}
\end{equation}
With that the Dirac equation is obtained as
\begin{equation}
\Pi^D (p) \Psi (p ) = \Psi (p)\, .
\label{eq_Dir}
\end{equation}
   
As next I consider the projector onto the $\lbrace {1\over 2}, {1\over 2}
\rbrace \otimes \lbrack \lbrace {1\over 2},0\rbrace \oplus \lbrace 0, 
{1\over 2} \rbrace \rbrack $ representation which is constructed as 
\cite{RumFet}
\begin{equation}
\Pi^{RS}_{\mu\nu} (p)  =  {1\over {2M}}(/\!\!\!p +M)(g_{\mu\nu} -
{{p_\mu p_\nu }\over M^2}\,)\, ,
\label{Rar_Schw}
\end{equation} 
where $M$ stands for the mass average of the resonance multiplet.
It acts on a 16--dimensional field $\Psi_\mu (p)$  that is mapped 
onto a 4--dimensional Lorentz vector with Dirac spinor components. 
A polar vector $\Psi_\mu (p) $ collects the D$_{2I\, ,3}$, S$_{2I\, ,1}$, 
and P$_{2I\, ,1}$ states of natural parities, whereas a pseudovector 
$\Psi_\mu (p) $ joins the P$_{2I\, ,3}$, P$_{2I\, ,1}$, and S$_{2I\, ,1}$ 
resonances. From Eq.~(\ref{Rar_Schw}) follows that the relativistic 
equation of motion for the first family of nucleon resonaces is
\begin{equation}
{1\over {2M}} (/\!\!\!p +M) 
(g^{\mu \nu } - { {p^\mu p^\nu }\over M^2}) \Psi_\mu (p)  = 
\Psi^\nu (p) \, , \label{res_degr}
\end{equation}
with $\Psi_\mu (p)  $ being a polar vector.
The exclusion of eight components from $\Psi_\mu (p)$ by means of
the subsidiary condition $\gamma^\mu\Psi_\mu (p)  = 0$ and Proca's 
equation, $\partial ^\mu \Psi_\mu (p) =0 $, is known as the 
Rarita--Schwinger equation \cite{Lurie} for a spin--${3\over 2}$ 
particle. It is well known to possess a pathological property. 
When  minimally coupled to an external electromagnetic field, the 
particle described by such an equation will move with superluminal 
velocity. This is exclusively due to the presence of the redundant 
components in the Rarita-Schwinger field $\Psi_\mu (p) $ that have 
to be eliminated by means of the subsidiary conditions mentioned above. 

{}From the above considerations follows that the redundant states
are in fact the S$_{2I, 1}$ and P$_{2I, 1}$ resonances.
Therefore, the 16--dimensional Rarita--Schwinger field 
in fact describes  the resonance family S$_{11}$, P$_{11}$,
and D$_{13}$ (or P$_{13}$) as a whole, rather than an isolated
spin--${3\over 2}$ state.
As long as the lowest $\Delta $ baryons are  isolated P$_{33}$ states, 
the Rarita--Schwinger equation
is not quite adequate for describing the $\Delta $(1232) and 
$\Delta $(1600) resonances. Rather, the latter have to be interpreted 
by means of the projector onto the $\lbrace {3\over 2},0\rbrace \otimes 
\lbrace 0, {3\over 2}\rbrace $ representation from 
Eq.~(\ref{lowest_Deltas}).

In general, relativistic equations containing time derivatives of higher 
than the first order have the same pathological properties \cite{Swieca}.
Because of that usage of Eq.~(\ref{Rar_Schw}) for resonance description
is in fact unrealistic. Rather, one has to look for alternative 
descriptions. Indeed, very recently, a new method for the relativistic 
description of $\lbrace j,0\rbrace \oplus \lbrace 0, j\rbrace $ states that 
avoids these difficulties was developed in Ref.~\cite{Ahluwalia}. 
There, the representations considered were treated as multicomponent
bi--vectors rather than as 2$j$--rank multispinors. Their covariant 
behavior with respect to Lorentz group transformation was ensured by 
construction in terms of the explicit representation of the boost operation 
within the corresponding 2(2$j$+1)--dimensional spaces. In this way the wave 
functions of spin-$j$ particles are mapped onto covariant objects regardless 
of any equation of motion. After that the field operators are introduced as 
superposition of wave packets of one--particle states of given
spin polarizations with the coefficients in front being just
the spinors constructed. Finally, propagators can be consistently defined 
by means of two-point correlation functions. The apparatus developed in 
Ref.~\cite{Ahluwalia} provides one with all the ingredients needed for 
treating at a phenomenological level intermediate excitations of spin-j 
particles in, say, photoproduction of mesons off 
the proton, thus opening a new field for theoretical activities.
{}From Eqs.~(\ref{eq_Dir}) and (\ref{res_degr}) follows 
further  that while the nucleon satisfies the 
single Dirac equation, the lowest S$_{11}$ state is among the solutions of 
Eq.~(\ref{res_degr}). In this sense the nucleon and the ${1\over 2}^-$ 
resonance satisfy different equations of motion.

To recapitulate, 
the multi--spinor representations of the Lorentz group in fact describe 
families of mass degenerate resonances of different spins and parities.
Only occasionally some of them turn out to possess equal spins and 
opposite parities without constituting parity doublets 
as they belong to the same O(4) multiplet and their orbital angular 
momenta differ by one unit.

\section{Summary}

To conclude I wish to emphasize that the symmetry properties of 
baryons with masses below (and around) 2 GeV in coordinate space are well 
understood in terms of SL(2,C) representations constituted of bi--spinors 
$\lbrace {1\over 2}, 0\rbrace \oplus \lbrace 0, {1\over 2}\rbrace $
coupled to O(4) multiplets of the type $\lbrace j,j\rbrace $ 
with $j$ half integer. The SL(2,C) baryon classification scheme predicts 
two new nucleon and three new $\Delta $ resonances. These are the
F$_{17}$ state near 1.7 GeV and a H$_{1, 11}$ state slightly above 
2 GeV, on the one hand, and the P$_{31}$, P$_{33}$, and D$_{33}$ 
resonances at $\sim $ 2.3 GeV, on the other hand.
All three nucleon resonances with masses below 1.6 GeV were shown to 
correspond to the $\lbrace {1\over 2}, {1\over 2} \rbrace $ multiplet 
filled with O(3) states of natural parity thus reflecting the
spontaneous selection of the positive parity for the vacuum
in the Nambu--Goldstone mode of chiral symmetry.
On the contrary, above 1.6 GeV all observed nucleon 
resonances were shown to be collected by O(4) multiplets 
filled with O(3) states of unnatural parity thus revealing the 
Wigner--Weyl mode of chiral symmetry. Threfore,   
in the transition between the nucleon and the second S$_{11}$ state
at 1650 MeV the chiral symmetry mode changes, and a
chiral phase transition is predicted to take place.
This observation might be relevant for RHIC experiments.

The appeal of the classification of the baryon excitations according to the 
representations of the group SL(2,C), the universal covering of the Lorentz
group, is mainly hidden in the possibility to identify the relativistic
equations of motion for baryons of higher spin with the projectors onto
the representations considered.
It was shown that the 16--dimensional Rarita-Schwinger field describes 
the family of S$_{2I, 1}$, $P_{2I, 1}$, and 
D$_{2I, 3}$ (or P$_{2I, 3}$) as a whole rather than, as widely used, an
isolated spin--${3\over 2}$ state.

\section{Acknowledgements}
I wish to thank Florian Scheck and Hartmuth Arenh\"ovel for the 
critical reading of the manuscript and several instructive remarks
as well as Siddhartha Sen and Andreas Wirzba for interest.

This work was supported by the Deutsche Forschungsgemeinschaft
(SFB 201) and partly by the School of Mathematics at the Trinity College 
Dublin, where the investigation was finished.

\begin{table}
\caption{ The unnatural parity 
$\lbrace {3\over 2}, {3\over 2}\rbrace \otimes
\lbrack \lbrace {1\over 2}, 0\rbrace \oplus 
\lbrace 0, {1\over 2}\rbrace \rbrack $ 
multiplets for N and $\Delta $ resonances.}
\begin{tabular}{rcll}
$N$(mass) & $I(J^\pi)$        & $ \Delta $ (mass) &$ I(J^\pi )$ \\
\hline
N(1650)     & ${1\over 2}({1\over 2}^-)$ & $ \Delta (1900) $       
&${3\over 2}({1\over 2}^-)$\\
N(1710)     & ${1\over 2}({1\over 2}^+)$ & $ \Delta (1910) $       & 
${3\over 2}({1\over 2}^+)$\\
N(1720)     & ${1\over 2}({3\over 2}^+)$ & $\Delta (1920)  $       & 
${3\over 2}({3\over 2}^+)$\\
N(1700)     & ${1\over 2}({3\over 2}^-)$ & $\Delta (1940)  $       & 
${3\over 2}({3\over 2}^-)$\\
N(1675)     & ${1\over 2}({5\over 2}^-)$ & $\Delta (1930)  $       & 
${3\over 2}({5\over 2}^-)$\\
N(1680)     & ${1\over 2}({5\over 2}^+)$ & $\Delta (1905)  $       & 
${3\over 2}({5\over 2}^+)$\\
predicted    & ${1\over 2}({7\over 2}^+)$ & $\Delta (1950)  $       
& ${3\over 2}({7\over 2}^+ )$
\end{tabular}
\end{table}   

\begin{table}
\caption{ The unnatural parity 
$\lbrace {5\over 2}, {5\over 2}\rbrace \otimes
\lbrack \lbrace {1\over 2}, 0\rbrace \oplus 
\lbrace 0, {1\over 2}\rbrace \rbrack $ 
multiplets for N and  $\Delta $ resonances.}
\begin{tabular}{rcll}
$N$(mass) & $I(J^\pi)$        & $ \Delta $ (mass) &$ I(J^\pi )$ \\
\hline
N(2090)     & ${1\over 2}({1\over 2}^-)$ & $ \Delta (2150) $       
&${3\over 2}({1\over 2}^-)$\\
N(2100)     & ${1\over 2}({1\over 2}^+)$ &  predicted        & 
${3\over 2}({1\over 2}^+)$\\
N(1900)     & ${1\over 2}({3\over 2}^+)$ &  predicted         & 
${3\over 2}({3\over 2}^+)$\\
N(2080)     & ${1\over 2}({3\over 2}^-)$ & predicted       & 
${3\over 2}({3\over 2}^-)$\\
N(2200)     & ${1\over 2}({5\over 2}^-)$ & $\Delta (2350)  $       & 
${3\over 2}({5\over 2}^-)$\\
N(2000)     & ${1\over 2}({5\over 2}^+)$ & $\Delta (2000)  $       & 
${3\over 2}({5\over 2}^+)$\\
N(1990)   & ${1\over 2}({7\over 2}^+)$ & $\Delta (2390)  $       
& ${3\over 2}({7\over 2}^+ )$\\
N(2190)   & ${1\over 2}({7\over 2}^-)$ & $\Delta (2200)  $       
& ${3\over 2}({7\over 2}^- )$\\
N(2250)   & ${1\over 2}({9\over 2}^-)$ & $\Delta (2400)  $       
& ${3\over 2}({9\over 2}^- )$\\
N(2220)   & ${1\over 2}({9\over 2}^+)$ & $\Delta (2300)  $       
& ${3\over 2}({9\over 2}^+ )$\\
predicted  & ${1\over 2}({{11}\over 2}^+)$ & $\Delta (2420)  $       
& ${3\over 2}({{11}\over 2}^+ )$
\end{tabular}
\end{table}   

\begin{figure}[htbp]
\centerline{\psfig{figure=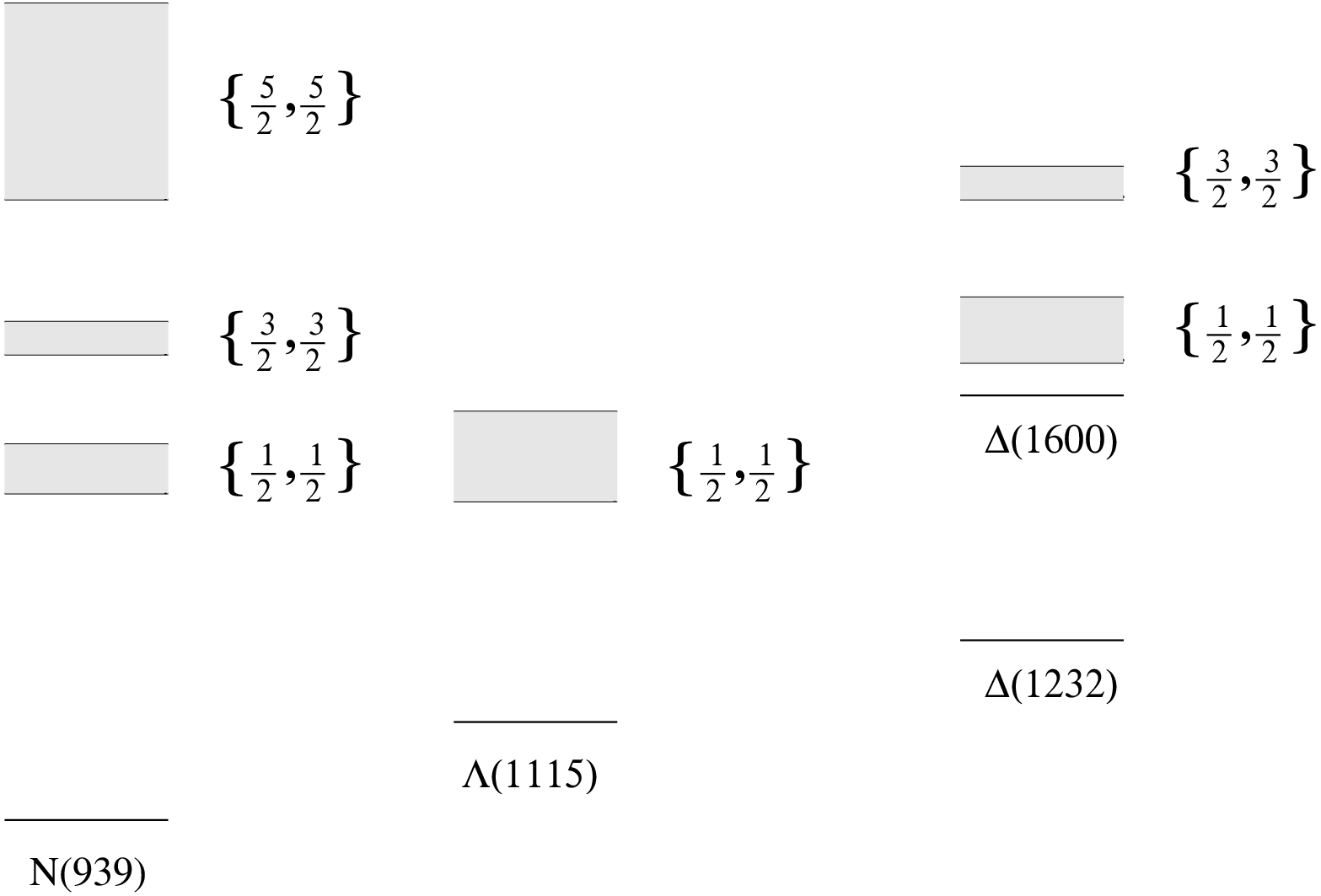,width=15cm}}
\vspace{0.1cm}
{\small Fig. 1\hspace{0.2cm} 
The gross structure of the baryon spectrum in terms of the
$\lbrace j,j\rbrace \otimes 
\lbrack \lbrace {1\over 2},0\rbrace \oplus \lbrace 0, {1\over 2}
\rbrace \rbrack $ representations of the SL(2,C) group. 
Only the $\lbrace j,j\rbrace $ 
assignements have been displayed. Each dashed area contains
$4j+1 $ resonances (compare text). 
 }
\end{figure}

\end{document}